\documentclass[runningheads]{llncs}
\usepackage{amssymb}
\usepackage[T1]{fontenc}
\usepackage{graphicx}
\usepackage{booktabs}
\usepackage{makecell}
\usepackage{verbatim}
\usepackage{enumitem}
\usepackage{float}
\usepackage{multirow} 
\makeatletter
\renewcommand*{\@fnsymbol}[1]{\ensuremath{\ifcase#1\or \dagger\or \dagger\dagger\or
   \mathsection\or \mathparagraph\or \|\or **\or \dagger\dagger
   \or \ddagger\ddagger \else\@ctrerr\fi}}
\makeatother

\begin{document}

\title{Moving from 2D to 3D: volumetric medical image classification for rectal cancer staging}
\titlerunning{Volumetric image classification for rectal cancer staging}

\author{Joohyung Lee\inst{1}\thanks{These authors contributed equally to this work.}\and
Jieun Oh\inst{2}$^{\dagger}$\and
Inkyu Shin\inst{1} \and
You-sung Kim\inst{3} \and
Dae Kyung Sohn\inst{4} \and
Tae-sung Kim\inst{2,3}\thanks{These co-corresponding authors contributed equally to this work.}  \and
In So Kweon\inst{1}$^{\dagger\dagger}$}

    

\institute{Korea Advanced Institute of Science and Technology, Daejeon, South Korea \email{iskweon77@kaist.ac.kr} \and
Healthcare AI Team, National Cancer Center, Goyang, South Korea \email{tsangel@ncc.re.kr}
\and Department of Radiology, National Cancer Center, Goyang, South Korea
\and Center for Colorectal Cancer, National Cancer Center, Goyang, South Korea}

\authorrunning{J. Lee et al.}

\date{February 2022}

\maketitle

\begin{abstract}
Volumetric images from Magnetic Resonance Imaging (MRI) provide invaluable information in preoperative staging of rectal cancer. Above all, accurate preoperative discrimination between T2 and T3 stages is arguably both the most challenging and clinically significant task for rectal cancer treatment, as chemo-radiotherapy is usually recommended to patients with T3 (or greater) stage cancer. In this study, we present a volumetric convolutional neural network to accurately discriminate T2 from T3 stage rectal cancer with rectal MR volumes. Specifically, we propose 1) a custom ResNet-based volume encoder that models the inter-slice relationship with late fusion (i.e., 3D convolution at the last layer), 2) a bilinear computation that aggregates the resulting features from the encoder to create a volume-wise feature, and 3) a joint minimization of triplet loss and focal loss. With MR volumes of pathologically confirmed T2/T3 rectal cancer, we perform extensive experiments to compare various designs within the framework of residual learning. As a result, our network achieves an AUC of 0.831, which is higher than the reported accuracy of the professional radiologist groups. We believe this method can be extended to other volume analysis tasks.

\keywords{Rectal Cancer Staging  \and Medical Volume Classification \and Convolutional Neural Network  \and Anisotropy.}

\end{abstract}


\section{Introduction}


Volumetric medical images (VMIs) are widely used image representation in medical field. For example, Magnetic Resonance Images (MRIs) play an essential role in treatment planning for patients as it enables preoperative T-staging \cite{trebeschi2017deep}. Specifically, the preoperative discrimination of T2 stage from T3 stage rectal cancer is arguably the most difficult yet clinically significant task since it determines whether to use chemo-radiotherapy, a physically burdensome and expensive treatment option~\cite{mri_of_rectal_cancer}. Despite the importance of accurate staging, both radiologists' and surgeons' ability to differentiate between the two stages using MRI vary widely~\cite{Tstage_1p5T}.

In our previous study~\cite{kim2019rectal}, we outlined a 2D convolutional neural network (CNN) model for T2/T3 discrimination that receives 2D MR images. However, it requires radiologists to manually select a representative slice (2D) from each MR volume (3D). In this study, we propose a 3D convolutional neural network (CNN) that classifies rectal MR volumes as T2 or T3 stages.

However, the decision to use either a 2D or 3D CNN for volume classification remains an open problem~\cite{nnunet}. Advocates for 2D convolution address the large anisotropy of volumetric images: for example, the x-y resolution of CT is more than ten times higher than that of its third axis \cite{liu20183d,peng2020saint}. Therefore, they claim that applying 3D CNNs with isotropic kernels to anisotropic medical volume can be problematic~\cite{anisotropy_DL_Techniques}. Nonetheless, 3D CNNs can aggregate information from the third axis and is therefore widely used in practice~\cite{3dunet,VNET}. In our work, we investigate ways to include both types of convolutions to create an anisotropic receptive field for VMI analysis.



To capture information from the third axis, most studies in VMI analysis utilize global average pooling~\cite{huang2021batch}. However, in fine-grained image recognition (FGIR) community, bilinear encoding is widely used as an aggregation function \cite{agg_bcnn}. Because VMI classification shares similarity with FGIR, bilinear encoding can be an effective depth aggregation function for VMI classification. Video classification is another area where studies have explored different aggregation functions \cite{agg_TSN}. Consequently, in this study, we explore various aggregation functions from the field of medical image analysis, FGIR, and video action recognition.

The main contributions of our work are as follows:

\begin{itemize}
\item To the best of our knowledge, this is the first reported system that automates rectal cancer staging with 3D MR volumes. 


\item We introduce a CNN model that classifies rectal MR volumes. Our encoder fuses the third axis information at the last encoding layer, which makes an anisotropic receptive field for an anisotropic MR volume. We also propose aggregating the resulting feature from the encoder using bilinear computation. We train the network with a joint minimization strategy of focal loss and triplet loss.

\end{itemize}


\section{Materials and Methods}
\begin{figure}[h]
\centering
\includegraphics[width=0.85\textwidth]{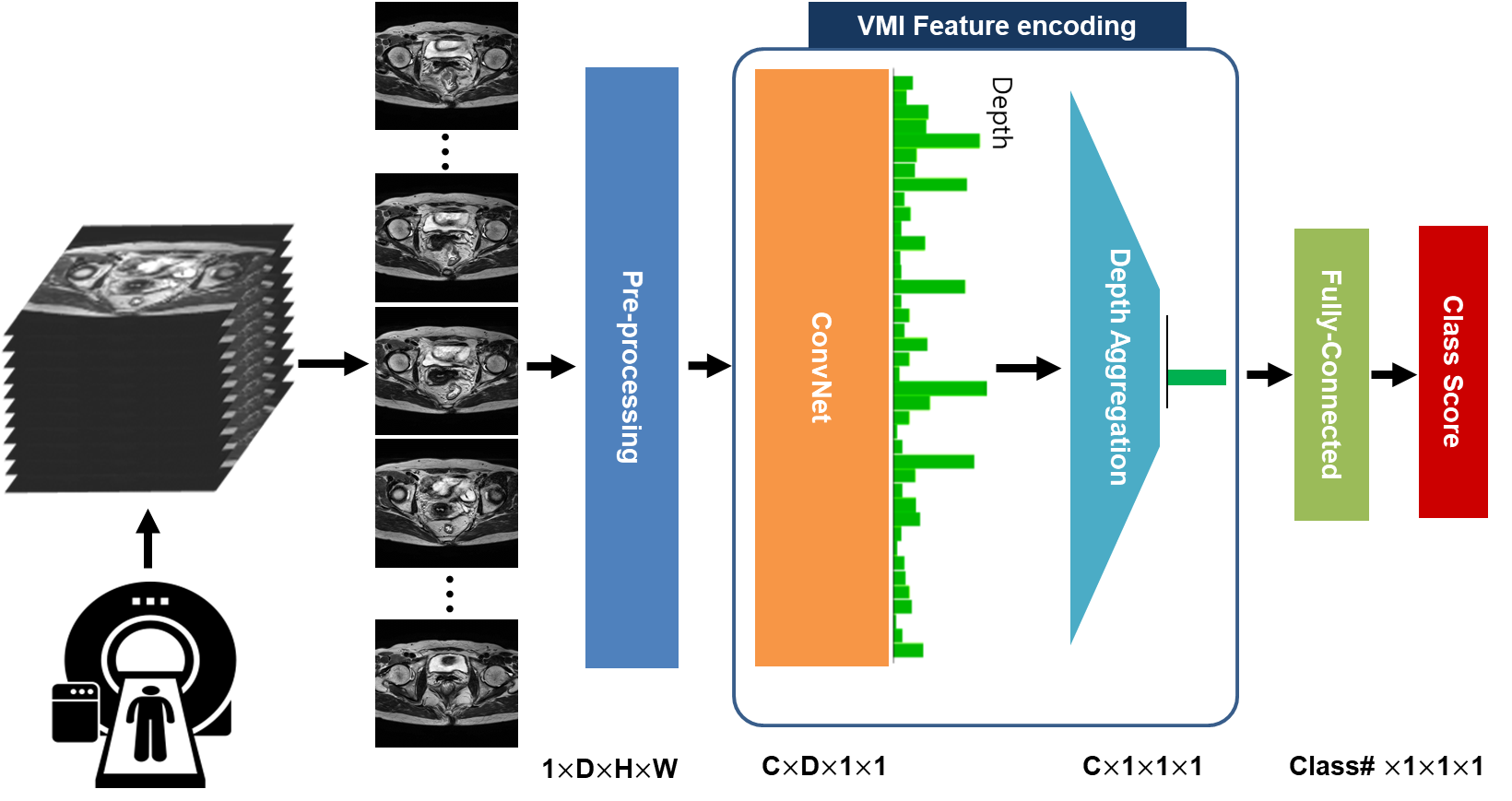}
\caption{Proposed volumetric medical image classification architecture. Our work focuses on both the network architecture and depth aggregation function.} \label{fig1}
\end{figure}

Our goal is to solve a clinically challenging problem: to distinguish T2-stage rectal cancer from T3-stage rectal cancer with MRI scans. To this end, we propose a volume classification model for anisotropic medical volumes (Fig.~\ref{fig1}). First, we aim to find an effective feature extractor. Second, we compare the performance of various aggregation functions that summarize the extracted features. Finally, we search for supplementary objective functions that improve model performance.

\subsection{Dataset and Preprocessing}
\label{section:dataset}
For this study, we retrospectively (2004-2018) collected 168 T2-stage and 399 T3-stage MR volumes from 567 patients with rectal cancer. The cancer stage labels were pathologically confirmed (determined at surgical resection). Scans were collected with one of seven 3.0T or 1.5T superconducting systems using pelvic phased-array coils. Table~\ref{tab:table_one} contains subject demographic characteristics and device information. For each MR volume, radiologists select one to three representative slices that best reflect the clinical state of the whole volume; we assign the pathologically confirmed cancer stage of an MR volume to all of its representative slice(s), which is our slice-level label. This study was conducted according to the principles of the Declaration of Helsinki, and the protocol was approved by the Institutional Review Board of our institution (NCC2019-0081).
\begin{table}[t]
    \caption{Subject demographic characteristics and device information. Discovery 750 3.0T, Genesis Signa 1.5T, Signa 1.5T, and Signa HDX 3.0T scanners were manufactured by GE Healthcare ($n=97$), Achieva 3.0T, Achieva TX 3.0T, and Ingenia CX 3.0T scanners by Philips Healthcare ($n=468$), and Skyra 3.0T scanners by Siemens ($n=2$). MFS: magnetic field strength.}\label{tab:table_one}
    \centering
    \begin{tabular}{cccc}
    \toprule
    {} &     All ($n=567$) &   T2 ($n=168$) &     T3 ($n=399$) \\
    \midrule
    Age               &     63.12$\pm$11.23 &  62.95$\pm$10.59 &    63.19$\pm$11.48 \\
    Sex (Female/Male) &         198/369 &       68/100 &        130/269 \\
    Tumor Volume (cm$^{3}$)   &       7.53$\pm$9.64 &    3.32$\pm$5.21 &      9.3$\pm$10.49 \\
    N Stage (0/1/2/x) &  259/155/133/20 &  120/27/7/14 &  139/128/126/6 \\
    MFS (1.5T/3T)     &          77/490 &       23/145 &         54/345 \\
    Slice Thickness (mm)  &        3.1$\pm$0.42 &     3.1$\pm$0.41 &      3.11$\pm$0.43 \\
    Pixel Spacing (mm)   &      0.439$\pm$0.069 &      0.443$\pm$0.069 &       0.437$\pm$0.069 \\

    \bottomrule
    \end{tabular}
\end{table}

For pre-processing, we apply N4ITK to correct the intensity inhomogeneity, following the conventions of MRI community \cite{n4itk}. Moreover, we conduct the following training-time augmentations: elastic transformation, optical distortion, grid distortion, random shift, random scale, random rotate, random crop, horizontal flip, CLAHE, random brightness, random contrast, motion blur, median blur, and Gaussian blur. Because localization around rectum region (rectum and tumor) is relatively simple task, and previously reported automatic methods have sufficiently high accuracy (around 94.3\% DSC) \cite{lee_ieee_access}, we start our experiment from the x-y localized setting (around the rectum and tumor) and resize along the first and second axes to create $256\times256\times N$ pixel volumes (where $N$ represents the number of slices in the third axis). We perform 10-fold cross-validation following the procedure of our previous study \cite{lee_ieee_access} and report the performance score of hold-out set as described in Section \ref{section:training_testing}. 


\begin{figure}[t]
\centering
\includegraphics[width=0.85\textwidth]{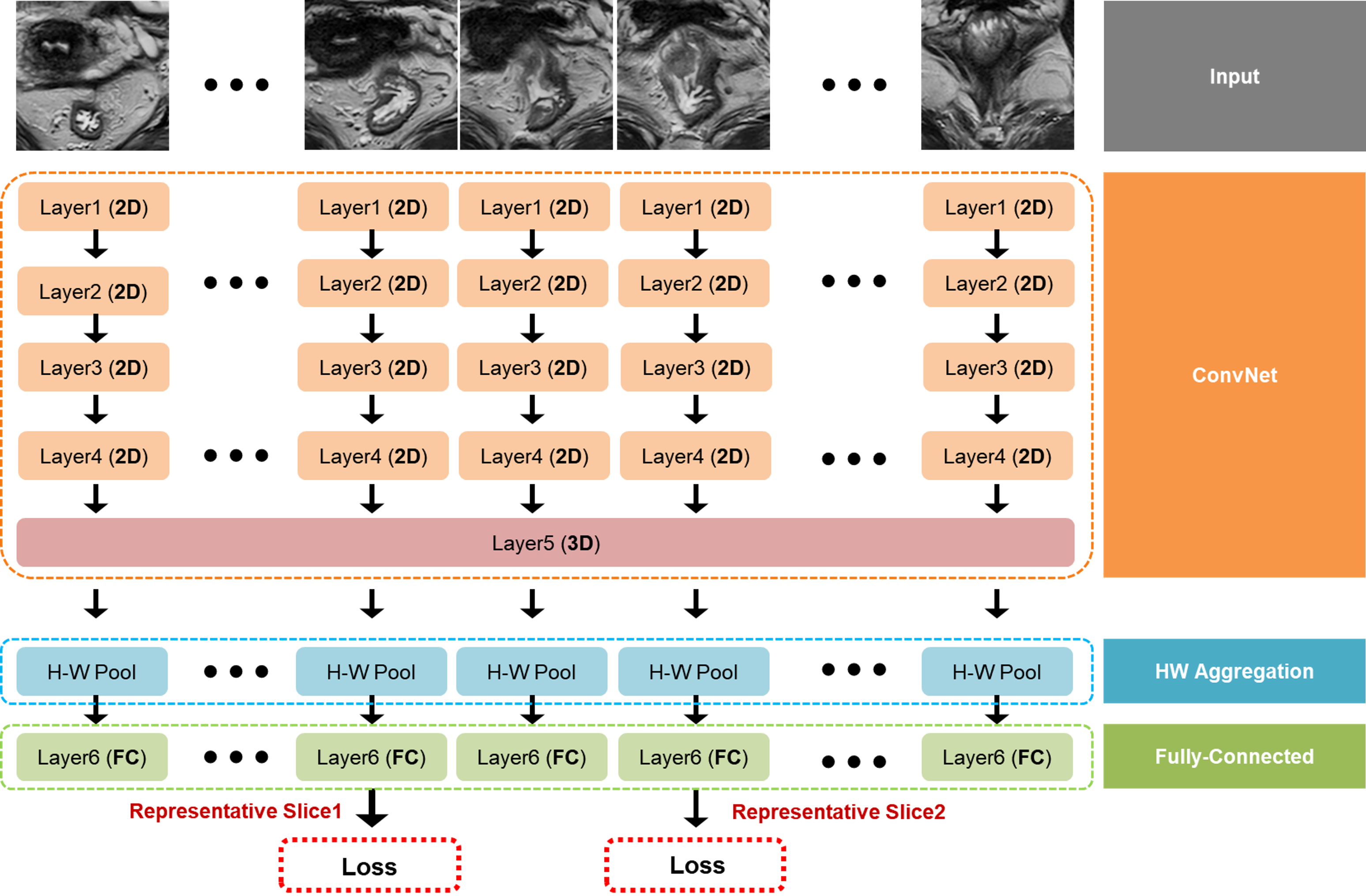}
\caption{Selected mixture($f$-$rMC_{5}$) of 2D CNN and 3D CNN to map volumetric medical image to slice-wise probability. } \label{fig2}
\end{figure}

\subsection{2D CNN vs. 3D CNN for Anisotropic Rectal Volume Analysis}

Our first experimental design compares various feature extractors for rectal MR volumes. Here, we address the debate on performance supremacy between 2D CNN and 3D CNN for anisotropic medical volumes. As 2D CNNs can only yield slice-wise scores, we compare the performance of 2D CNN and 3D CNN in slice-level classification by using slice-level labels as described in Section~\ref{section:dataset}. Therefore, our feature extractor is $f : \mathbb{R}^{H\times W\times D} \rightarrow \mathbb{R}^{D\times 256}$ as illustrated in Fig.~\ref{fig2}. This experimental design evaluates whether adjacent slices help slice-wise classification. Motivated by Tran \textit{et al.} \cite{video_closer}, we also compare the performance of various mixtures of 2D and 3D convolutions($f$-$MC_{x}$, $f$-$rMC_{x}$), which create anisotropic receptive fields. However, unlike Tran \textit{et al.}, we did not stride along the third axis because our goal is to yield slice-wise scores and to examine if training with adjacent slices improves performance. Models with names starting with `f-' are models that yield slice-wise scores.
The mixtures of 2D and 3D convolution consist of two types: $f$-$MC_{x}$ and $f$-$rMC_{x}$. $f$-$MC_{x}$ fuse the third axis information in the early layer and thus have 3D convolution in the early layer(s), i.e., 1st to (x-1)th layer, and have 2D convolution in the late layer, i.e., xth layer to the last layer, whereas $f$-$rMC_{x}$ is the opposite. All of our CNN architectures are based on ResNet-18 with the 32 filters for the first convolution layer~\cite{resnet}. To train our feature extractors, we map the slice-wise features to slice-wise probabilities through a fully-connected layer that is shared across all slices. From the outputted slice-wise cancer stage probabilities, we choose only the probabilities for the representative slices to train the feature extractor as illustrated in Fig.~\ref{fig2}.

\subsection{Supplementary Objective Function}

We train 2D CNN, 3D CNN, and mixed 2D and 3D CNN candidates using focal loss~\cite{focal_loss} for our feature extractor experiments. This loss function is a generalization of cross-entropy loss with an additional down-weighting parameter for when the prediction is close to the ground truth. The loss can be expressed as follows, where $p$ is the softmax probability:

\begin{equation}\label{eq:eq_loss_focal}
\mathcal{L}_{focal}(p)=-(1-p)^\gamma\log(p)
\end{equation}







Once we determine the best performing residual block (see Section~\ref{section:2Dand3D}), we compare the performance of the various supplementary loss functions. These functions are added to focal loss and trained jointly to enhance inter-class separability and intra-class compactness~\cite{weinberger2009distance}. Specifically, we implement center loss \cite{centerloss} and triplet loss \cite{triplet}. We use the suggested hyperparameters settings from their respective original manuscripts. The losses are defined as follows:

\begin{equation}\label{eq:eq_loss_center}
\mathcal{L}_{center}=\frac{1}{2} \sum_{i=1}^{m} \| x_i-c_{y_i} \|_2^2
\end{equation}

\begin{equation}\label{eq:eq_loss_triplet}
    \mathcal{L}_{triplet}(x, x^+, x^-; f)=\\max(0,\|f-f^+\|_2^2-\|f-f^-\|_2^2+m)
\end{equation}

The \(y_i\)th class center of deep features is denoted by \(c_{y_i} \in \mathbb{R}^{c'} \). To select both the positive and negative samples for triplet loss, we implement online-hard mining from Schroff \textit{et al.} \cite{triplet}.

\subsection{Depth Aggregation Function}

Global average pooling is a common choice for aggregation functions in CNNs~\cite{resnet,mobileNet,video_closer}. However, as proper classification of medical volumes is usually contingent on a couple of slices (representative slices), global average pooling may not be the optimal solution. Using the best performing objective function investigated in Section~\ref{section:objective_function}, we evaluate the performance of different aggregation functions on our rectal MR dataset. With the number of extracted feature dimension $C$, and the number of slices of input VMI $D$, our depth aggregation function is defined as $f : \mathbb{R}^{C\times D} \rightarrow \mathbb{R}^{C\times 1}$. Specifically, we experiment with four different types of aggregation functions: average pooling, max pooling, attention weighting \cite{agg_TSN}, and bilinear encoding \cite{agg_bcnn}. When $i$ and $d$ are the channel index and depth dimension respectively, attention weighting is defined as follows:
\begin{equation}\label{eq:depthAGG_attention1}
g_{i, att}=\sum_{d=1}^{D}A(i,d)f_i^d
\end{equation}
To design the attention weighting function $A(i,d)$, we did not create an additional embedding space but rather directly applied a softmax function. Therefore, the attention weight vector becomes as follows:

\begin{equation}\label{eq:depthAGG_attention1}
A(i,d)=\frac{exp(f_i^d)}{\sum_{d'=1}^{D}exp(f_i^{d'})}
\end{equation}
Functions other than attention weighting are straightforward. Max-pooling and average-pooling select the maximum and average value along depth per each channel dimension. For bilinear encoding, we used the same features twice to generate the bilinear feature with $l$\textsubscript{2} normalization scheme \cite{agg_bcnn}.

\subsection{Training and Testing}
\label{section:training_testing}
To mitigate data imbalance between T2 and T3 stage, we perform oversampling. We train all models using Stochastic Gradient Descent(SGD) with initial learning rate of 0.01 and applied weight decay with ratio($\lambda$) 0.01 to mitigate overfitting. Moreover, to report the final predictions performance, we follow the test-time augmentation procedure from Wu \textit{et al.} \cite{tmi_breast}. As a performance metrics, we report mean and standard deviation of average area-under-curve (AUC), accuracy, sensitivity, and specificity over ten folds. We select the best performing module and loss based on mean AUC.

\begin{table}[h]
\centering
\caption{Performance comparison of different mixture of 2D CNN and 3D CNN.}\label{tab1}
\begin{tabular}{cccccc}
\hline
\bfseries CNN & \bfseries AUC & \bfseries Acc & \bfseries Recall(T2) & \bfseries Recall(T3) & \bfseries \# params\\
\hline
    $f$-$R2D$ & 0.798$\pm$0.042 & 75.0$\pm$2.9 & 60.0$\pm$12.6 & 81.4$\pm$7.0 & 2796001 \\
  $f$-$R3D$ & 0.763$\pm$0.074 & 74.5$\pm$5.3 & 55.5$\pm$12.9 & 82.5$\pm$7.0 & 8291873\\
  $f$-$R(2+1)D$ & 0.758$\pm$0.069 & 74.3$\pm$5 & 50.1$\pm$12.8 & 84.5$\pm$5.6 & 8294563\\
  $f$-$MC_{2}$ & 0.789$\pm$0.061 & 73.5$\pm$3.8 & 65.5$\pm$13.3 & 76.9$\pm$5.1 & 2799137\\
  $f$-$MC_{3}$ & 0.746$\pm$0.07 & 70.2$\pm$9.4 & 58.8$\pm$16.0 & 74.9$\pm$17.1 & 2872865\\
  $f$-$MC_{4}$ & 0.767$\pm$0.059 & 72.5$\pm$3.5 & 55.7$\pm$12.8 & 79.7$\pm$6.5 & 3130913\\
  $f$-$MC_{5}$ & 0.737$\pm$0.042 & 72.3$\pm$4.8 & 49.7$\pm$8.8 & 81.9$\pm$8.0 & 4163105\\
  $f$-$rMC_{2}$ & 0.778$\pm$0.051 & 75.1$\pm$4.3 & 55.6$\pm$13.9 & 83.3$\pm$6.5 & 8288737\\
  $f$-$rMC_{3}$ & 0.796$\pm$0.052 & 75.1$\pm$2.2 & 59.0$\pm$15.3 & 81.8$\pm$6.8 & 8215009\\
  $f$-$rMC_{4}$ & 0.787$\pm$0.055 & 76.2$\pm$4.5 & 50.0$\pm$13.6 & 87.3$\pm$4.6 & 7956961\\
  $f$-$rMC_{5}$ & \textbf{0.815}$\pm$\textbf{0.057} & 77.3$\pm$4.4 & 48.5$\pm$10.3 & 89.4$\pm$5.0 & 6924769\\
\hline
\end{tabular}
\end{table}

\section{Results}

\subsection{2D CNN vs. 3D CNN for Anisotropic Rectal Volume Analysis}
\label{section:2Dand3D}
Table~\ref{tab1} contains the performance metrics for models with different mixtures of 2D and 3D convolutions. First, the results show that 2D CNN($f$-$R2D$) outperforms 3D CNN($f$-$R3D$) for our task. Moreover, $f$-$rMC_{x}$ tends to excel $f$-$MC_{x}$, which shows that late fusion($f$-$rMC_{x}$) along the third axis is more effective to model inter-slice relationship than early fusion($f$-$MC_{x}$). Among all the models, $f$-$rMC_{5}$, which is illustrated in Fig.~\ref{fig2}, scores the highest AUC followed by $f$-$R2D$; adjacent slices do help learning their center slice. Specifically, the only difference between $f$-$rMC_{5}$ and $f$-$R2D$ is that $f$-$rMC_{5}$ makes inter-slice modeling at the last layer of the network. Moreover, $f$-$R(2+1)D$ doesn't excel unlike in video classification task. We attribute this to the difference between video and medical volume; the third axis of medical volume is spatial as the other two axes whereas that of video is temporal. It has to be noted that $f$-$R2D$ yields no receptive field along the third axis whereas the receptive field of $f$-$R3D$ is elongated towards the third axis because the pixel spacing along the third axis is much larger than that along the first and the second.

\subsection{Supplementary Objective Function}
\label{section:objective_function}
Table~\ref{tab2} compares the classification performance of different supplementary loss functions. When combined with focal loss, center loss lowers performance whereas the triplet loss boosts the performance of the baseline, i.e., focal loss. Specifically, by adding center loss to focal loss, the performance dropped by 0.012 in AUC. Adding triplet loss to focal loss increases the AUC by 0.019. The addition of the focal loss and triplet loss will be used for all following experiments in this study.

\begin{table}[b]
\centering
\caption{Performance comparison among supplementary loss functions.}\label{tab2}
\begin{tabular}{ccccc}
\hline
  \bfseries Loss & \bfseries AUC & \bfseries Acc & \bfseries Recall(T2) & \bfseries Recall(T3)\\
\hline
  Focal & 0.815$\pm$0.057 & 77.3$\pm$4.4 & 48.5$\pm$10.3 & 89.4$\pm$5.0 \\
  +Center & 0.803$\pm$0.041 & 74.9$\pm$3.9 & 49.$\pm$13.9 & 85.8$\pm$5.4 \\
  +Triplet & \textbf{0.834}$\pm$ \textbf{0.068} & 78$\pm$4.9 & 49.9$\pm$9.7 & 89.8$\pm$4.7\\
\hline
  \end{tabular}
\end{table}

\subsection{Evaluation on Aggregation Functions}

\begin{table}[h]
\centering
\caption{Performance comparison of depth aggregation functions to summarize depth-wise feature along depth-axis.}\label{tab3}
\begin{tabular}{ccccc}
\hline
\bfseries Pool & \bfseries AUC & \bfseries Acc & \bfseries Recall(T2) & \bfseries Recall(T3)\\
\hline
  AVP & 0.821$\pm$0.055 & 76$\pm$3.8 & 34$\pm$9.6 & 93.7$\pm$4.9\\
  MXP & 0.815$\pm$0.047 & 78.1$\pm$4.1 & 47$\pm$12.6 & 91.2$\pm$5.9\\
  Bilin & \textbf{0.831}$\pm$ \textbf{0.062} & 79.3$\pm$6.2 & 62.1$\pm$13.5 & 86.7$\pm$6.8\\
  Att & 0.817$\pm$0.057 & 77.2$\pm$4.8 & 48.3$\pm$14.7 & 89.5$\pm$4.2\\
\hline
  \end{tabular}
\end{table}

With CNN module ($f$-$rMC_{5}$) and loss function (focal loss with triplet loss) selected, the performances of various depth-aggregation functions are evaluated and compared. Note that we have already aggregated x-y information by striding and by average pooling as depicted in Fig.~\ref{fig2}. The outcome of x-y pooling, which is the slice-level feature, is then summarized into a single volume-level feature by the depth aggregation function. As a result, selecting average pooling as a depth aggregation function equals applying global average pooling after the last convolution layer.
Here we have four candidates, including the relatively basic: 1) average pooling, 2) max pooling, and the more complex: 3) bilinear encoding, and 4) attention weighting. Attention weighting is implemented with the hope that attention mechanism may be able to find the representative slices from the volume. No additional parameters are introduced for depth aggregation function except that the feature space is enlarged by bilinear encoding. The results are described in Table~\ref{tab3}.

As shown in Table~\ref{tab3}, bilinear encoder scores the best performance among four functions to summarize the depth information. Because bilinear encoder has been successful in the field of FGIR, we assume that the ability of bilinear encoder to capture fine-grained detail of tumour can be successful in cancer staging as well \cite{agg_bcnn,zheng2019learning}. Moreover, bilinear encoder notably closes the gap between Acc(T2) and Acc(T3). Note that the performance degradation from the triplet loss of Table~\ref{tab2} and average pooling of Table~\ref{tab3} shows that volume level classification is more challenging than slice-level classification. It may be because volume has more unnecessary areas than a single slice in staging the cancer level.

\subsection{Performance Comparison with Radiologists}
Two studies have reported the performance of professional radiologists in discriminating T2-stage and T3-stage rectal cancer. Since the datasets are different, direct comparison is not possible. However, performance can still be compared to earn some idea about how competitive our method is (Table~\ref{tab4}).

\begin{table}[h]
\centering
\caption{Discrimination performance of T2/T3 rectal cancer by radiologists (Ang \textit{et al.}, Maas et al \textit{et al.}) and by our proposed method}
\label{tab4}
\begin{tabular}{ccccc}
\hline
\bfseries Method & \bfseries AUC & \bfseries Acc & \bfseries Recall(T2) & \bfseries Recall(T3)\\
\hline
  \makecell{Ang \textit{et al.} \cite{accuracy_pelvicMRI}} & - & 68.87 & 85.5 & 50.1\\
  \makecell{Maas \textit{et al.} (1.5T MR) \cite{Tstage_1p5T} } & 0.73 & 66.67 & 83.33 & 52.38\\
  \makecell{Maas \textit{et al.} (3.0T MR) \cite{Tstage_1p5T} } & 0.64 & 56.41 & 72.22 & 42.86\\
  \makecell{$f$-$rMC_{5}$+Bilinear+Triplet (Ours)} & \textbf{0.831} & 79.3& 62.1& 86.7\\
  \hline
  \end{tabular}
\end{table}


\section{Conclusion}
This paper proposes a CNN model to preoperatively discriminate between T2 and T3 stage rectal cancer from rectal MR volumes. Our network consists of two parts: a CNN feature extractor and a depth aggregation function. The CNN maps medical volumes to slice-wise features while the aggregation function summarizes the slice-wise features into a volume-wise feature. To investigate the best performing CNN for anisotropic rectal VMI, we compared the performance with varying approach to model inter-slice relationship. As a result, $f$-$rMC_{5}$, which represents the late inter-slice interaction, is selected. Moreover, through extensive experimentation, we found that using triplet loss and bilinear encoder as a depth aggregation function and supplementary objective function, respectively, outperformed other functions and losses. To the best of our knowledge, this is the first study to apply the experimental approach from the video action community to address the debate on whether 3D convolutions are an efficient method for anisotropic VMI models. We believe our proposed CNN can be used for other anisotropic VMI-related tasks such as VMI segmentation and detection. 


\subsubsection{Acknowledgments}
The authors would like to thank Young Sang Choi for his helpful feedback. This work was supported by KAIST R\&D Program (KI Meta-Convergence Program) 2020 through Korea Advanced Institute of Science and Technology (KAIST), a grant from the National Cancer Center (NCC2010310-1), and the National Research Foundation of Korea(NRF) grant funded by the Korea government(MSIT) (NRF-2020R1C1C1012905).

\bibliographystyle{splncs04}

\bibliography{bibliography}

\begin{thebibliography}{10}
\providecommand{\url}[1]{\texttt{#1}}
\providecommand{\urlprefix}{URL }
\providecommand{\doi}[1]{https://doi.org/#1}

\bibitem{accuracy_pelvicMRI}
Ang, Z.H., De~Robles, M.S., Kang, S., Winn, R.: Accuracy of pelvic magnetic
  resonance imaging in local staging for rectal cancer: a single local health
  district, real world experience. ANZ Journal of Surgery  \textbf{91}(1-2),
  111--116 (2021)

\bibitem{3dunet}
{\c{C}}i{\c{c}}ek, {\"O}., Abdulkadir, A., Lienkamp, S.S., Brox, T.,
  Ronneberger, O.: 3d u-net: learning dense volumetric segmentation from sparse
  annotation. In: International conference on medical image computing and
  computer-assisted intervention. pp. 424--432. Springer (2016)

\bibitem{resnet}
He, K., Zhang, X., Ren, S., Sun, J.: Deep residual learning for image
  recognition. In: Proceedings of the IEEE conference on computer vision and
  pattern recognition. pp. 770--778 (2016)

\bibitem{anisotropy_DL_Techniques}
Hesamian, M.H., Jia, W., He, X., Kennedy, P.: Deep learning techniques for
  medical image segmentation: achievements and challenges. Journal of digital
  imaging  \textbf{32}(4),  582--596 (2019)

\bibitem{mri_of_rectal_cancer}
Horvat, N., Carlos Tavares~Rocha, C., Clemente~Oliveira, B., Petkovska, I.,
  Gollub, M.J.: Mri of rectal cancer: Tumor staging, imaging techniques, and
  management. Radiographics  \textbf{39}(2),  367--387 (2019)

\bibitem{mobileNet}
Howard, A., Sandler, M., Chu, G., Chen, L.C., Chen, B., Tan, M., Wang, W., Zhu,
  Y., Pang, R., Vasudevan, V., et~al.: Searching for mobilenetv3. In:
  Proceedings of the IEEE/CVF International Conference on Computer Vision. pp.
  1314--1324 (2019)

\bibitem{huang2021batch}
Huang, Z., Zhou, Q., Zhu, X., Zhang, X.: Batch similarity based triplet loss
  assembled into light-weighted convolutional neural networks for medical image
  classification. Sensors  \textbf{21}(3), ~764 (2021)

\bibitem{nnunet}
Isensee, F., Petersen, J., Klein, A., Zimmerer, D., Jaeger, P.F., Kohl, S.,
  Wasserthal, J., Koehler, G., Norajitra, T., Wirkert, S., et~al.: nnu-net:
  Self-adapting framework for u-net-based medical image segmentation. arXiv
  preprint arXiv:1809.10486  (2018)

\bibitem{kim2019rectal}
Kim, J., Oh, J.E., Lee, J., Kim, M.J., Hur, B.Y., Sohn, D.K., Lee, B.: Rectal
  cancer: Toward fully automatic discrimination of t2 and t3 rectal cancers
  using deep convolutional neural network. International Journal of Imaging
  Systems and Technology  \textbf{29}(3),  247--259 (2019)

\bibitem{lee_ieee_access}
Lee, J., Oh, J.E., Kim, M.J., Hur, B.Y., Sohn, D.K.: Reducing the model
  variance of a rectal cancer segmentation network. IEEE Access  \textbf{7},
  182725--182733 (2019)

\bibitem{focal_loss}
Lin, T.Y., Goyal, P., Girshick, R., He, K., Doll{\'a}r, P.: Focal loss for
  dense object detection. In: Proceedings of the IEEE international conference
  on computer vision. pp. 2980--2988 (2017)

\bibitem{agg_bcnn}
Lin, T.Y., RoyChowdhury, A., Maji, S.: Bilinear convolutional neural networks
  for fine-grained visual recognition. IEEE transactions on pattern analysis
  and machine intelligence  \textbf{40}(6),  1309--1322 (2017)

\bibitem{liu20183d}
Liu, S., Xu, D., Zhou, S.K., Pauly, O., Grbic, S., Mertelmeier, T., Wicklein,
  J., Jerebko, A., Cai, W., Comaniciu, D.: 3d anisotropic hybrid network:
  Transferring convolutional features from 2d images to 3d anisotropic volumes.
  In: International Conference on Medical Image Computing and Computer-Assisted
  Intervention. pp. 851--858. Springer (2018)

\bibitem{Tstage_1p5T}
Maas, M., Lambregts, D.M., Lahaye, M.J., Beets, G.L., Backes, W., Vliegen,
  R.F., Osinga-de Jong, M., Wildberger, J.E., Beets-Tan, R.G.: T-staging of
  rectal cancer: accuracy of 3.0 tesla mri compared with 1.5 tesla. Abdominal
  imaging  \textbf{37}(3),  475--481 (2012)

\bibitem{VNET}
Milletari, F., Navab, N., Ahmadi, S.A.: V-net: Fully convolutional neural
  networks for volumetric medical image segmentation. In: 2016 fourth
  international conference on 3D vision (3DV). pp. 565--571. IEEE (2016)

\bibitem{peng2020saint}
Peng, C., Lin, W.A., Liao, H., Chellappa, R., Zhou, S.K.: Saint: spatially
  aware interpolation network for medical slice synthesis. In: Proceedings of
  the IEEE/CVF Conference on Computer Vision and Pattern Recognition. pp.
  7750--7759 (2020)

\bibitem{triplet}
Schroff, F., Kalenichenko, D., Philbin, J.: Facenet: A unified embedding for
  face recognition and clustering. In: Proceedings of the IEEE conference on
  computer vision and pattern recognition. pp. 815--823 (2015)

\bibitem{video_closer}
Tran, D., Wang, H., Torresani, L., Ray, J., LeCun, Y., Paluri, M.: A closer
  look at spatiotemporal convolutions for action recognition. In: Proceedings
  of the IEEE conference on Computer Vision and Pattern Recognition. pp.
  6450--6459 (2018)

\bibitem{trebeschi2017deep}
Trebeschi, S., van Griethuysen, J.J., Lambregts, D.M., Lahaye, M.J., Parmar,
  C., Bakers, F.C., Peters, N.H., Beets-Tan, R.G., Aerts, H.J.: Deep learning
  for fully-automated localization and segmentation of rectal cancer on
  multiparametric mr. Scientific reports  \textbf{7}(1), ~1--9 (2017)

\bibitem{n4itk}
Tustison, N.J., Avants, B.B., Cook, P.A., Zheng, Y., Egan, A., Yushkevich,
  P.A., Gee, J.C.: N4itk: improved n3 bias correction. IEEE transactions on
  medical imaging  \textbf{29}(6),  1310--1320 (2010)

\bibitem{agg_TSN}
Wang, L., Xiong, Y., Wang, Z., Qiao, Y., Lin, D., Tang, X., Van~Gool, L.:
  Temporal segment networks for action recognition in videos. IEEE transactions
  on pattern analysis and machine intelligence  \textbf{41}(11),  2740--2755
  (2018)

\bibitem{weinberger2009distance}
Weinberger, K.Q., Saul, L.K.: Distance metric learning for large margin nearest
  neighbor classification. Journal of machine learning research  \textbf{10}(2)
  (2009)

\bibitem{centerloss}
Wen, Y., Zhang, K., Li, Z., Qiao, Y.: A discriminative feature learning
  approach for deep face recognition. In: European conference on computer
  vision. pp. 499--515. Springer (2016)

\bibitem{tmi_breast}
Wu, N., Phang, J., Park, J., Shen, Y., Huang, Z., Zorin, M., Jastrz{\k{e}}bski,
  S., F{\'e}vry, T., Katsnelson, J., Kim, E., et~al.: Deep neural networks
  improve radiologists’ performance in breast cancer screening. IEEE
  transactions on medical imaging  \textbf{39}(4),  1184--1194 (2019)

\bibitem{zheng2019learning}
Zheng, H., Fu, J., Zha, Z.J., Luo, J.: Learning deep bilinear transformation
  for fine-grained image representation. Advances in Neural Information
  Processing Systems  \textbf{32} (2019)

\end{thebibliography}

\end{document}